\documentclass[useAMS,usenatbib]{mn2e}
\def\jnl@style{\it}
\def\aaref@jnl#1{{\jnl@style#1}}

\def\aaref@jnl#1{{\jnl@style#1}}

\def\aj{\aaref@jnl{AJ}}                   
\def\araa{\aaref@jnl{ARA\&A}}             
\def\apj{\aaref@jnl{ApJ}}                 
\def\apjl{\aaref@jnl{ApJ}}                
\def\apjs{\aaref@jnl{ApJS}}               
\def\ao{\aaref@jnl{Appl.~Opt.}}           
\def\apss{\aaref@jnl{Ap\&SS}}             
\def\aap{\aaref@jnl{A\&A}}                
\def\aapr{\aaref@jnl{A\&A~Rev.}}          
\def\aaps{\aaref@jnl{A\&AS}}              
\def\azh{\aaref@jnl{AZh}}                 
\def\baas{\aaref@jnl{BAAS}}               
\def\jrasc{\aaref@jnl{JRASC}}             
\def\memras{\aaref@jnl{MmRAS}}            
\def\mnras{\aaref@jnl{MNRAS}}             
\def\pra{\aaref@jnl{Phys.~Rev.~A}}        
\def\prb{\aaref@jnl{Phys.~Rev.~B}}        
\def\prc{\aaref@jnl{Phys.~Rev.~C}}        
\def\prd{\aaref@jnl{Phys.~Rev.~D}}        
\def\pre{\aaref@jnl{Phys.~Rev.~E}}        
\def\prl{\aaref@jnl{Phys.~Rev.~Lett.}}    
\def\pasp{\aaref@jnl{PASP}}               
\def\pasj{\aaref@jnl{PASJ}}               
\def\qjras{\aaref@jnl{QJRAS}}             
\def\skytel{\aaref@jnl{S\&T}}             
\def\solphys{\aaref@jnl{Sol.~Phys.}}      
\def\sovast{\aaref@jnl{Soviet~Ast.}}      
\def\ssr{\aaref@jnl{Space~Sci.~Rev.}}     
\def\zap{\aaref@jnl{ZAp}}                 
\def\nat{\aaref@jnl{Nature}}              
\def\iaucirc{\aaref@jnl{IAU~Circ.}}       
\def\aplett{\aaref@jnl{Astrophys.~Lett.}} 
\def\apspr{\aaref@jnl{Astrophys.~Space~Phys.~Res.}}
\def\bain{\aaref@jnl{Bull.~Astron.~Inst.~Netherlands}} 
\def\fcp{\aaref@jnl{Fund.~Cosmic~Phys.}}  
\def\gca{\aaref@jnl{Geochim.~Cosmochim.~Acta}}   
\def\grl{\aaref@jnl{Geophys.~Res.~Lett.}} 
\def\jcp{\aaref@jnl{J.~Chem.~Phys.}}      
\def\jgr{\aaref@jnl{J.~Geophys.~Res.}}    
\def\jqsrt{\aaref@jnl{J.~Quant.~Spec.~Radiat.~Transf.}}
\def\memsai{\aaref@jnl{Mem.~Soc.~Astron.~Italiana}}
\def\nphysa{\aaref@jnl{Nucl.~Phys.~A}}   
\def\physrep{\aaref@jnl{Phys.~Rep.}}   
\def\physscr{\aaref@jnl{Phys.~Scr}}   
\def\planss{\aaref@jnl{Planet.~Space~Sci.}}   
\def\procspie{\aaref@jnl{Proc.~SPIE}}   

\usepackage{graphicx}
\usepackage{amsfonts}
\usepackage{amssymb}
\usepackage{amsmath}
\usepackage{gensymb}
\usepackage{fixltx2e}
\usepackage{color}
\usepackage[caption = false]{subfig}
\title[Multiscale Analysis of the Gradient of $\bmath{P}$]{Multiscale Analysis of the Gradient of Linear Polarisation}

\author[J.-F. Robitaille and A. M. M. Scaife]{J.-F. Robitaille$^{1}$\thanks{E-mail:
jean-francois.robitaille@manchester.ac.uk; anna.scaife@manchester.ac.uk} and A. M. M. Scaife$^{1}$\footnotemark[1]\\
$^{1}$Jodrell Bank Centre for Astrophysics, School of Physics and Astronomy,
\\ The University of Manchester, Oxford Road, Manchester M13 9PL, UK}

\begin{document}

\date{to be submitted}

\pagerange{\pageref{firstpage}--\pageref{lastpage}} \pubyear{2015}

\maketitle

\label{firstpage}

\begin{abstract}

We propose a new multiscale method to calculate the amplitude of the gradient of the linear polarisation vector, $|\nabla \bmath{P}|$,  using a wavelet-based formalism. We demonstrate this method using a field of the Canadian Galactic Plane Survey (CGPS) and show that the filamentary structure typically seen in $|\nabla \bmath{P}|$ maps depends strongly on the instrumental resolution. Our analysis reveals that  different networks of filaments are present on different angular scales. The wavelet formalism allows us to calculate the power spectrum of the fluctuations seen in $|\nabla \bmath{P}|$ and to determine the scaling behaviour of this quantity. The power spectrum is found to follow a power law with $\gamma \approx 2.1$. We identify a small drop in power between scales of $80 \lesssim l \lesssim 300$ arcmin, which corresponds well to the overlap in the $u$--$v$ plane between the Effelsberg 100--m telescope and the DRAO 26--m telescope data. We suggest that this drop is due to undersampling present in the 26--m telescope data. In addition, the wavelet coefficient distributions show higher skewness on smaller scales than at larger scales. The spatial distribution of the outliers in the tails of these distributions creates a coherent subset of filaments correlated across multiple scales, which trace the sharpest changes in the polarisation vector $\bmath{P}$ within the field. We suggest that these structures may be associated with highly compressive shocks in the medium. The power spectrum of the field excluding these outliers shows a steeper power law with $\gamma \approx 2.5$.

\end{abstract}


\section{Introduction}

Previous power spectrum analysis of the amplitude of the polarisation intensity vector, $|\bmath{P}|= \sqrt{Q^2+U^2}$, where $Q$ and $U$ are the Stokes parameters, in the Galactic plane has shown evidence of large-scale structures in the Galactic magnetic field \citep{2003A&A...403.1045H, 2014ApJ...787...34S}. The power-law behaviour of these spectra is expected to be related to the energy transfer from larger to smaller scales in the turbulent fluctuations of the magnetic field. Power-law variations as a function of Galactic latitude have also been measured \citep{2003A&A...403.1045H}, as well as localised variations in regions associated with HII regions or supernova remnants \citep{2014ApJ...787...34S}. Unfortunately, the power spectrum of $|\bmath{P}|$ alone is not sensitive to fluctuations of the polarisation angle, $\theta=(1/2)\arctan(U/Q)$, which also show evidence of large-scale variations in the Galactic plane. According to \citet{2010A&A...520A..80L}, large-scale variations of $\theta$ are probably associated with the large-scale features of the magnetic field aligned with the spiral structure of the Galaxy. On the other hand, fluctuations of $\theta$ on smaller scales could be explained by a turbulent Faraday screen in front of a uniform polarised background. However, such perfect conditions are almost never satisfied in the interstellar medium (ISM) and most fluctuations seen in Stokes $Q$ and $U$ are probably due to a combination of Faraday rotation and intervening polarised emission along the line of sight in the Galactic plane. For these reasons, the interpretation of the direction and the amplitude of $\bmath{P}$ when considered separately is very difficult. 

\citet{2011Natur.478..214G} proposed calculation of the amplitude of the gradient of $\bmath{P}$, $|\nabla \bmath{P}|$, as a new technique to measure variations of the vector $\bmath{P}$ in the $Q$--$U$ plane. It is defined as,

\begin{equation}
|\nabla \bmath{P}| = \sqrt{ \left(\frac{\partial Q}{\partial x}\right)^2 +  \left(\frac{\partial U}{\partial x}\right)^2 +  \left(\frac{\partial Q}{\partial y}\right)^2  + \left(\frac{\partial U}{\partial y}\right)^2 },
\label{eq:gradient_P}
\end{equation}

\noindent This quantity can trace changes in both the direction and the amplitude of the vector $\bmath{P}$. Acting as an edge detector in a map, the gradient of $\bmath{P}$ highlights areas of sharp change in the magnetic field and/or the free-electron density, which are most likely due to turbulent fluctuations or shock fronts in the ISM. Since $|\nabla\bmath{P}|$ is only sensitive to the smallest scales, it is not significantly affected by the loss of large-scale structure in interferometric data. On the other hand, one disadvantage of using the gradient is that it may enhance noise present in the data and the distribution of its amplitude depends on the telescope resolution \citep{2012ApJ...749..145B}.

Variations of the emission probability distribution function (PDF) width as a function of angular resolution and angular scale were measured in thermal dust emission, infrared emission as well as in $|\nabla \bmath{P}|$ structures \citep{2010MNRAS.406.1350F, 2012ApJ...749..145B, 2014MNRAS.440.2726R}. In general, this may be explained by small-scale high-extinction cores or structures that are not present on larger scales and do not create the large skewness typical of the lognormal distribution usually measured in star formation regions. The PDF of gas, dust column density and $|\nabla \bmath{P}|$ is also expected to reflect the signature of physical processes occurring in the medium, e.g. turbulence, gravitational collapse or shocks \citep{2011MNRAS.416.1436B, 2012ApJ...749..145B, 2013ApJ...766L..17S}. All those physical processes are scale dependent: shocks usually produce fine scale structures, gravitational collapse depends on the local density of the gas and turbulence has the ability to transfer energy from large to smaller scales. Since the free-electron density and the magnetic field can also be affected by shocks and turbulence in the ISM, fluctuations in the polarisation intensity traced by $|\nabla\bmath{P}|$ should also be present on a broad range of scales. Previous studies using $|\nabla\bmath{P}|$ maps \citep{2011Natur.478..214G, 2012ApJ...749..145B, 2014A&A...566A...5I} concentrated their analysis on small-scale fluctuations, primarily for two reasons: (1) the gradient of a map only samples the smallest scales and (2) many radio observations are made interferometrically and miss information on large-scale structures since they do not completely sample the Fourier plane. In this paper, we propose a method to generalise such $|\nabla\bmath{P}|$ analysis to multiple scales using data where single-dish measurements are present.

The calculation of the gradient as an edge detector in two-dimensional images has found multiple applications in different fields. \citet{Canny1986} has shown, with image analysis methods for computer vision, that there is an uncertainty principle related to the detection and the localisation of a noisy step edge. In the presence of low signal to noise data, the precision of the localisation of edges must be traded by applying the gradient to a Gaussian-smoothed image. \citet{Canny1986} also show that the first derivative of Gaussians of different width can be used directly as a multiscale edge detector. In a similar vein, \citet{Mallat1992} generalised the method using wavelet transforms in a singularity detection algorithm applied to one and two-dimensional signals. Later, this generalised method, called the wavelet transform modulus maxima (WTMM) was used by \citet{Arneodo2000} as a multifractal analysis tool.

In this work, we propose a similar technique based on wavelet analysis to calculate $|\nabla \bmath{P}|$ as a function of scale for data where single-dish measurements are present (a description of these data is presented in  Section \ref{sec:observation}). A description of the resolution effect on the calculation of $|\nabla \bmath{P}|$ is presented in Section \ref{sec:resolution}; the wavelet formalism is presented in Section \ref{sec:DoG}; the formalism is tested on simulations in Section \ref{sec:test}; application to real data and discussion are presented in Sections \ref{sec:results} and \ref{sec:discussion} and conclusions in Section \ref{sec:conclusion}.

\section{Observations}\label{sec:observation}

\begin{figure}
\centering
\includegraphics[]{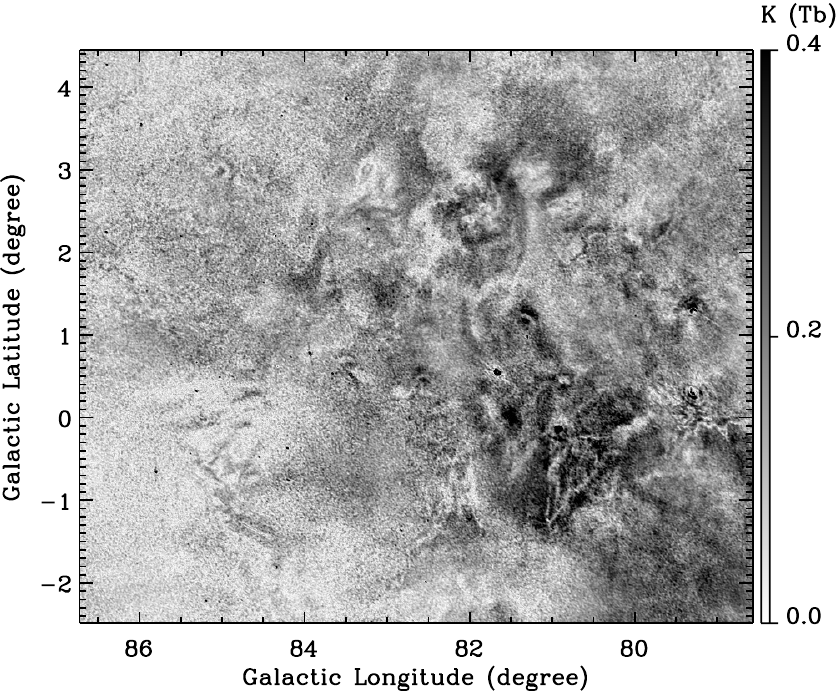}
\caption{The amplitude of the polarisation intensity vector, $|\bmath{P}|$, calculated from Stokes $Q$ and $U$ maps of the CGPS data.\label{fig:MM12_MN12_P}}
\end{figure}

The analysis in this work is applied to a field of the Canadian Galactic Plane Survey (CGPS; \citealt{2003AJ....125.3145T}) at 1420 MHz on polarised data including Stokes parameters $Q$ and $U$. For this survey, interferometric data were observed with the Synthesis Telescope at the Dominion Radio Astrophysical Observatory (DRAO; \citealt{2000A&AS..145..509L}). All observations were then completed with lower spatial frequencies from the Effelsberg 100--m telescope and the DRAO 26--m Telescope \citep{2003AJ....125.3145T, 2010A&A...520A..80L}. The chosen field is a combination of four mosaics from the survey available at the Canadian Astronomy Data Centre (CADC)\footnote{http://www1.cadc-ccda.hia-iha.nrc-cnrc.gc.ca}. It covers $\sim 8$\degree\, in Galactic longitude and $\sim 7$\degree\, in Galactic latitude. The field is centred at $l=82.65$\degree\, and $b=0.98$\degree. The amplitude of the polarisation intensity vector, $|\bmath{P}|$, is shown in Fig. \ref{fig:MM12_MN12_P}. CGPS data have an angular resolution of $\sim1 \csc \delta$ arcmin (where $\delta$ is the declination) and a pixel size of 18 arcsec.

\section{Resolution effect on $|\nabla \bmath{P}|$} \label{sec:resolution}

One advantage of the calculation of $|\nabla \bmath{P}|$ is that the results are not significantly affected by missing large-scale structures in interferometric data not completed with single-dish measurements. However, by sampling only the smallest scales, the gradient may enhance the noise in the data \citep{2012ApJ...749..145B}. Given the angular resolution of the CGPS maps and their pixel size, the synthesised beam is over-sampled by a factor of $\sim3.3$ \citep{2003AJ....125.3145T}, which means that, for these maps, the gradient is sensitive to variations smaller than the synthesised beam.

\begin{figure}
\centering
\includegraphics[]{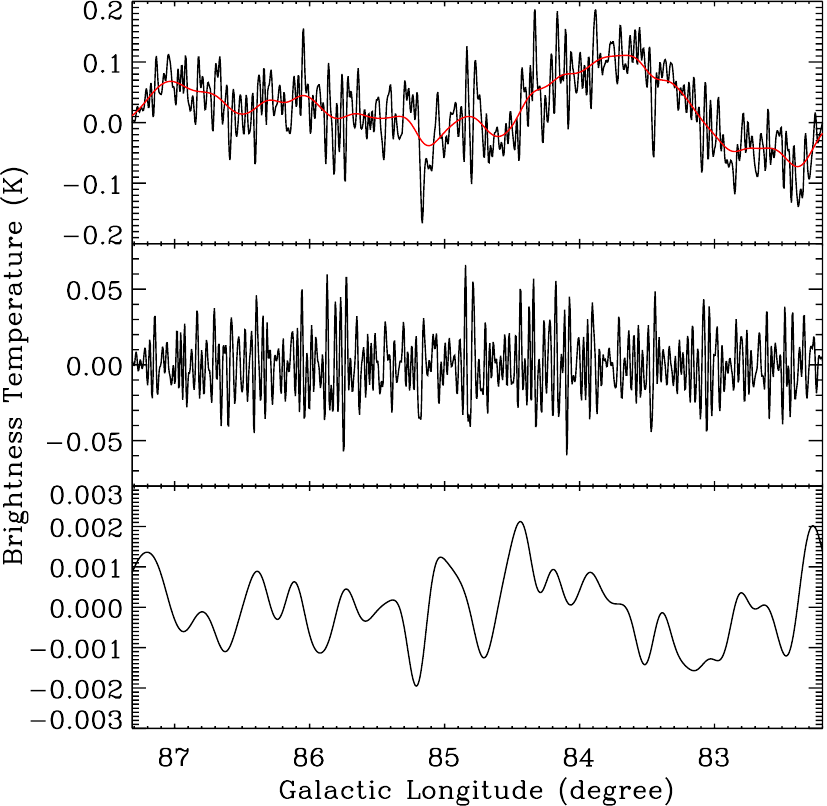}
\caption{A one dimensional signal representing one row of pixels located at $l=-0.72$\degree\, from the CGPS Stokes $Q$ image. From the top to the bottom are shown the original signal with a superposed smoothed version with a Gaussian filter having a standard deviation of $2^5$ pixels (red line), the first derivative of the original signal and the first derivative of the smoothed version.\label{fig:1d_deriv}}
\end{figure}

To visualise the resolution effect on the first derivative of a signal, the derivative of a one-dimensional function and its smoothed counterpart are shown in Fig. \ref{fig:1d_deriv}. The signal represents one row of pixels at $l=-0.72$\degree\, from the CGPS Stokes $Q$ image. To create a smoothed version of the signal, a Gaussian filter with a standard deviation of $2^5$ pixels is convolved with the original signal. The smoothed counterpart is shown superposed on the original signal in the top panel of Fig. \ref{fig:1d_deriv}. In the second panel, we show that in spite of obvious variations on larger scales in the one-dimensional signal, the first derivative of the original signal is much more sensitive to the smallest scale variation. On the other hand, the first derivative of the smoothed signal highlights variations on larger scales which are independent of variations seen at smaller scale.

\begin{figure*}
\centering
\includegraphics[]{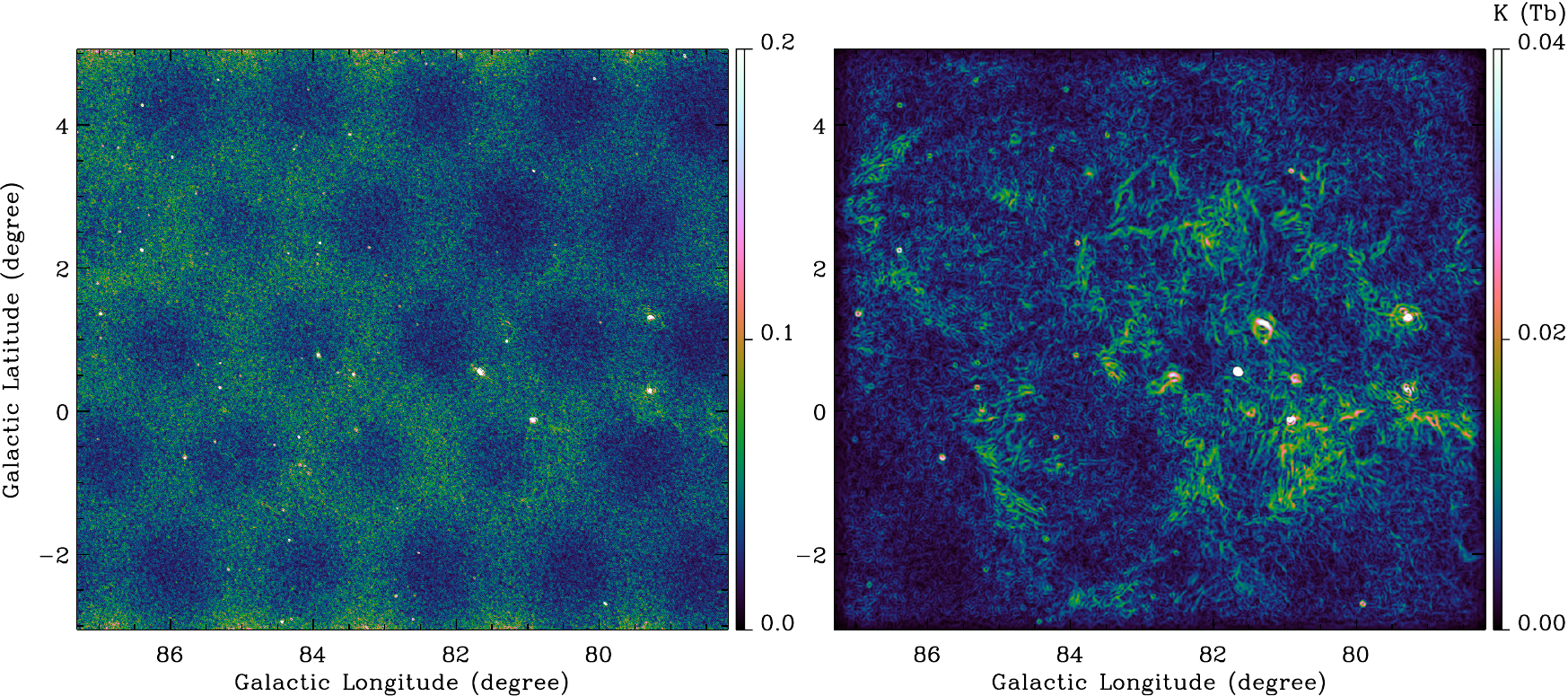}
\caption{The spatial gradient of linearly polarized emission, $|\nabla \bmath{P}|$, at the original resolution (left) and for the smoothed Stokes Q and U maps. Smoothed maps are produced with a convoluted Gaussian filter having a standard deviation of $2^2$ pixels.}
\label{fig:deltaP_smoothed}
\end{figure*}

Figure \ref{fig:deltaP_smoothed} shows the spatial gradient of the linearly polarized emission, $|\nabla \bmath{P}|$, for the CGPS field. In the left panel, filamentary structures normally identified from the calculation of $|\nabla \bmath{P}|$ can hardly be distinguished from variations at small scales associated with noise. However, the ``honeycomb'' noise variation pattern caused by the  survey mapmaking becomes clearly visible. The right panel shows the gradient calculated from Q and U maps smoothed using a Gaussian beam with a standard deviation of 4 pixels. Since the first derivative is only sensitive to fluctuations larger than the synthesised beam, filamentary structures, similar to those initially presented by \citet{2011Natur.478..214G}, are now clearly visible.

The effect of the map resolution on PDF moments of $|\nabla \bmath{P}|$ values was previously observed by \citet{2012ApJ...749..145B}. They used a Gaussian smoothing with FWHMs from 3 to 9 pixels on $Q$ and $U$ maps and measured significant decreases in all of the first four moments of $|\nabla \bmath{P}|$ for simulations of turbulence with supersonic Mach numbers. Figure \ref{fig:1d_deriv} and \ref{fig:deltaP_smoothed} illustrate that smoothing maps of $Q$ and $U$ not only changes the distribution of $|\nabla \bmath{P}|$ values but can also significantly change its structures. In the next section, we propose a multiscale analysis technique in order to visualise and quantify changes in the distribution of $|\nabla \bmath{P}|$ maps as a function of scale.

\section{DoG wavelet analysis} \label{sec:DoG}
\subsection{Formalism} \label{subsec:formalism}

The convolution of a Gaussian beam with maps of $Q$ and $U$ before the calculation of their gradient can give access to variations and sharp changes of $|\nabla \bmath{P}|$ at different angular resolution. By applying the technique on multiple angular scales, i.e. by gradually changing the Gaussian beam width convolved with maps $Q$ and $U$, it is possible to extend the analysis of $|\nabla \bmath{P}|$ images in the spatial frequency domain.

Following the work of \citet{Canny1986} and \citet{Mallat1992}, wavelet transforms can be used as a basis for developing a multiscale edge detector analysis. The wavelet transform of a signal consists of the convolution of a set of functions, called daughter wavelets, each of which represents a scaled version of a mother wavelet. One class of wavelet functions called the Derivative of Gaussian (DoG) is defined by

\begin{equation}
\psi(\bmath{x}) = (-1)^m \frac{\textrm{d} ^m}{\textrm{d}|\bmath{x}|^m} \phi(\bmath{x}),
\label{eq:mother_DoG}
\end{equation}

\noindent where

\begin{equation}
\begin{array}{rl}

\phi(\bmath{x})& = \frac{1}{2\pi} e^{\frac{-|\bmath{x}|^2}{2}}\\

			& = \frac{1}{2\pi } e^{\frac{-(x^2+y^2)}{2}}.\\

\end{array}
\label{eq:Gaussian}
\end{equation}

\noindent The second order ($m=2$) DoG wavelet represents the widely used ``Mexican Hat'' continuous wavelet. Even values of $m$ create symmetric functions which are appropriate for most general applications of wavelet transforms. Odd values of $m$ create asymmetric functions which are useful for revealing directional trends in data. They can also be used as edge detectors for structures present in an image. For the purpose of this analysis, the order of the mother wavelet will take the value of 1 or 3.

In this section, polarised data are considered as two-dimensional functions $f(\bmath{x})$, where $\bmath{x}$ is the vector position in a $x$--$y$ plane. The continuous wavelet transform of $f(\bmath{x})$ with the DoG wavelet can be expressed as

\begin{equation}
\tilde{f}(l,\bmath{x}) =
\begin{cases}
\tilde{f}_1 = l^{-2} \int \psi_1[l^{-1}(\bmath{x'}-\bmath{x})]f(\bmath{x}) d^2\bmath{x'}\\
\tilde{f}_2 = l^{-2} \int \psi_2[l^{-1}(\bmath{x'}-\bmath{x})]f(\bmath{x}) d^2\bmath{x'},
\end{cases}
\label{eq:DoG_transforms}
\end{equation}

\begin{figure*}
\centering
\includegraphics[]{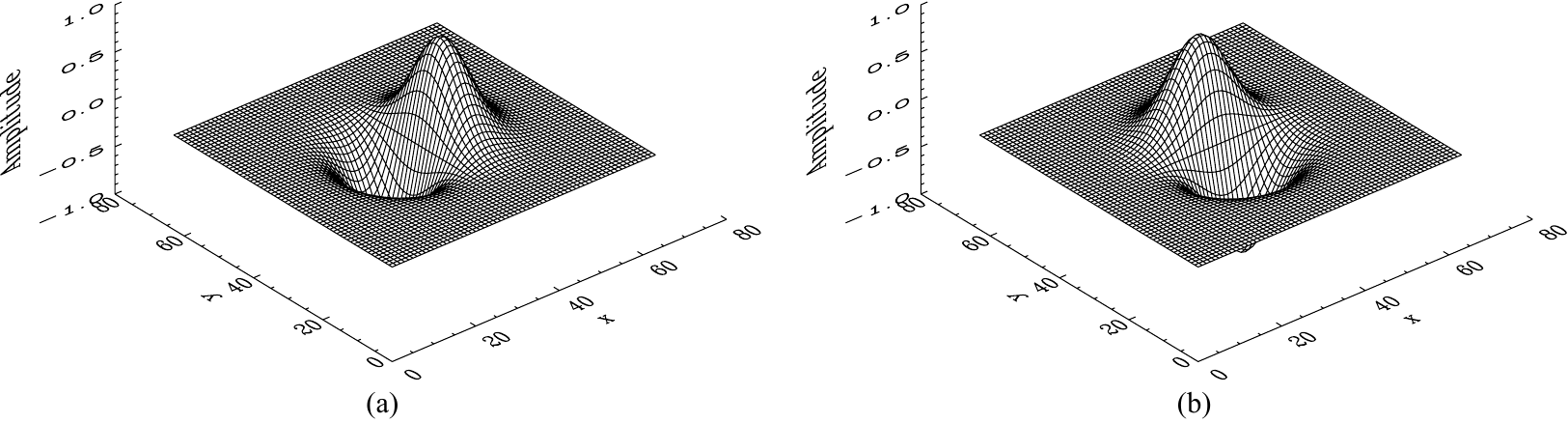}
\caption{Analysing wavelets (a) $-\psi_1$ and (b) $-\psi_2$. (The negative sign is for a better visualisation of the functions.)}
\label{fig:DoG}
\end{figure*}

\noindent where $\psi_1(x,y) = \partial^m \phi(x,y)/ \partial x^m$ and $\psi_2(x,y) = \partial^m \phi(x,y)/ \partial y^m$ (a three dimensional representation of these functions for $m=1$ is shown in Fig. \ref{fig:DoG}). All convolutions can be computed in the Fourier domain, which increases the speed of calculation. The function $\tilde{f}$ represents the wavelet transform of $f$ and $l$, the scaling factor of the wavelet. For $m=1$, it is interesting to note that equation (\ref{eq:DoG_transforms}) is equivalent to the calculation of the gradient of $f(\bmath{x})$ smoothed by dilated versions of a Gaussian beam:

\begin{equation}
\tilde{f}(l,\bmath{x}) = \nabla \{\phi(l^{-1}\bmath{x}) \otimes f(\bmath{x}) \},
\label{eq:grad_Gauss}
\end{equation}

\noindent where $\otimes$ is the convolution operation. From this point of view, the wavelet transform gives us a useful mathematical formalism on which a multiscaled version of $|\nabla \bmath{P}|$ can be defined. According to the statements above, $|\nabla \tilde{P}(l,\bmath{x})|$ can now be defined as:

\begin{equation}
|\nabla \tilde{P}(l,\bmath{x})| = \sqrt{ |\tilde{Q}(l,\bmath{x})|^2 + |\tilde{U}(l,\bmath{x})|^2},
\label{eq:scaled_gradient_P}
\end{equation}

\noindent where, referring to equation (\ref{eq:DoG_transforms}),

\begin{equation}
\begin{array}{rl}

|\tilde{Q}(l,\bmath{x})| & = \sqrt{ |\tilde{Q}_{1}(l,\bmath{x})|^2 + |\tilde{Q}_{2}(l,\bmath{x})|^2},\\
\\
|\tilde{U}(l,\bmath{x})| & = \sqrt{ |\tilde{U}_{1}(l,\bmath{x})|^2 + |\tilde{U}_{2}(l,\bmath{x})|^2}.\\

\end{array}
\label{eq:Stokes_amplitude}
\end{equation}

Since the work of \citet{Arneodo2000}, the mathematical formalism described by equations (\ref{eq:mother_DoG}) to (\ref{eq:grad_Gauss}) has been usually associated with the WTMM method. In order to extend the $|\nabla \bmath{P}|$ analysis to multiple angular scales, here we apply some components of this method to the CGPS polarization maps, in conjunction with a number of complementary methods inspired by the wavelet analysis techniques, as the $\Delta$-variance. However, a complete multifractal analysis, the original motivation for WTMM methods, is beyond the scope of this paper.

\subsection{Maxima chains}\label{subsec:maxima_chains}

Visually, the most interesting regions in maps of $|\nabla \bmath{P}|$ are areas showing the sharpest changes of the polarisation vector $\bmath{P}$. According to the WTMM method, one easy way to highlight these regions is by calculating the ``maxima chains'' of modulus values of the gradient.

In addition to the magnitude of the polarisation gradient, one can also calculate the argument or the direction of $\nabla \bmath{P}$, at each position in a map \citep{2011Natur.478..214G}:

\begin{equation}
\begin{array}{l}
\arg(\nabla \bmath{P})\equiv \\
\tan^{-1}\left (C_{\textrm{sign}} \sqrt{ \left ( \frac{\partial Q}{\partial y} \right )^2 + \left ( \frac{\partial U}{\partial y} \right )^2 } \bigg / \sqrt{ \left ( \frac{\partial Q}{\partial x} \right )^2 + \left ( \frac{\partial U}{\partial x} \right )^2 } \right ),\\
\end{array}
\label{eq:argument}
\end{equation}

\noindent where,

\begin{equation}
C_{\textrm{sign}}=\textrm{sign} \left ( \frac{\partial Q}{\partial x} \frac{\partial Q}{\partial y} +  \frac{\partial U}{\partial x} \frac{\partial U}{\partial y} \right ).
\label{eq:sign}
\end{equation}

\noindent Following this definition, modulus maxima are positions where $|\nabla \bmath{P}|$ is locally maximum in the direction of $\arg(\nabla \bmath{P})$. Thus, for every pixel on all scales, the argument of $\nabla \bmath{P}$ is calculated and the associated magnitude $|\nabla \bmath{P}|$ is compared with adjacent pixels having a similar $\arg(\nabla \bmath{P})$ value:  orientations are divided into only six different directions to take into account the pixelisation effect. After an iterative process for the entire map, maxima positions should lie on connected ``maxima chains''. Those chains allow us to visualise locations where strong fluctuations in the electron density distribution and/or magnetic field strength occur. Chains are also useful to visualise coherent structures that are ``connected'' through multiple scales.

\subsection{Wavelet power sprectrum}\label{sec:wav_pow}

Similarly to the Fourier transform, the wavelet transform conserves the total energy of the original signal. This property can be defined following the generalisation of the Plancherel identity for the continuous wavelet transform \citep{1992AnRFM..24..395F, 2011AJ....141...41D}:

\begin{equation}
\int |f(\bmath{x})|^2 d^2\bmath{x} = C_{\psi}^{-1}\int\int \frac{|\tilde{f}(l,\bmath{x})|^2}{l^2} dl d^2\bmath{x},
\label{eq:Plancherel}
\end{equation}

\noindent where

\begin{equation}
C_{\psi}=\int \frac{|\hat{\psi}(\bmath{k})|^2}{|\bmath{k}|^2}d^2\bmath{k} < \infty.
\label{eq:admissibility}
\end{equation}

\noindent Equation (\ref{eq:admissibility}) is also called the admissibility condition of the wavelet, where $\hat{\psi}(\bmath{k})$ is the Fourier transform of the mother wavelet $\psi(\bmath{x})$ and $\bmath{k}$ is the wavenumber vector. This condition is satisfied for every $m$-th order of equation (\ref{eq:mother_DoG}). From equation (\ref{eq:Plancherel}), the energy conservation can also be defined as a function of spatial scale only:

\begin{equation}
E(l)= \int \frac{|\tilde{f}(l,\bmath{x})|^2}{l^2} d^2\bmath{x}.
\label{eq:wavelet_energy}
\end{equation}

\noindent This relation shows that wavelet coefficients can be compared to Fourier coefficients and, as for the calculation of the Fourier power spectrum, wavelet coefficients can be used to measure the energy transfer from large to smaller scales. It is important to note that the normalisation factor $l^{-2}$ in equation (\ref{eq:DoG_transforms}) is only required to ensure the validity of equation (\ref{eq:grad_Gauss}) \citep{Arneodo2000}. In order to calculate the wavelet power spectrum of $|\nabla \tilde{P}(l,\bmath{x})|$, the regular normalisation of a wavelet transform, $l^{-1}$, is used. The wavelet energy spectrum defined by equation (\ref{eq:wavelet_energy}) can also be expressed in terms of the Fourier energy spectrum, $E(\bmath{k})=|\hat{f}(\bmath{k})|^2$ \citep{1992AnRFM..24..395F}:

\begin{equation}
E(l)= \int E(\bmath{k}) |\hat{\psi}(l\bmath{k})|^2 d^2\bmath{k}.
\label{eq:Fourier_wavelet}
\end{equation}

\noindent This relation means that at a particular scale, the global wavelet energy corresponds to the integral of the Fourier energy spectrum of the analysed function weighted with the energy spectrum of the wavelet at that scale.

In order to produce a wavelet power spectrum similar to the classical Fourier power spectrum, which takes into account the finite size of map $f(\bmath{x})$ and the discrete number of pixels, we use the relation:

\begin{equation}
S_P(l)= \frac{1}{N_xN_y} \sum_{\bmath{x}} |\nabla \tilde{P}(l,\bmath{x})|^2.
\label{eq:wavelet_power}
\end{equation}

\noindent The notation $S_P(l)$ is used here instead of the usual $P(l)$ for the power spectrum, in order to avoid possible confusion with the polarisation intensity $P$.

\subsection{Equivalence with Fourier wavelength}\label{sec:Fourier_equiv}

If one wants to compare the wavelet scale $l$ with the wavenumber $k$ in the Fourier domain, the equivalence of the scaling factor $l$ in the frequency domain has to be defined. The wavelet analysis described in the previous sections is equivalent to the calculation of the gradient of an image smoothed by Gaussian filters of different widths (see equation (\ref{eq:grad_Gauss})). Following this statement, the scale $l$ defined in the previous sections is related to the standard deviation of the Gaussian filter. For the following analysis, the wavelet scaling factor will be defined as $l_{\textrm{F}} = l\cdot(2\pi)^{-1}$, so that the scale $l$ can be directly compared to the wavenumber $k=1/l$ in the Fourier domain.

The $m$th derivative of the Gaussian makes the function oscillate around zero. The wavelength of these oscillations and their amplitude, rapidly decaying towards infinity, are the two properties which allow the wavelet function to be localised in the frequency domain. The width of the function in the frequency domain acts as a bandpass filter. According to \citet{2005CG.....31..846K}, one easy way to define the relationship between the scale of the wavelet function and the frequency content of the signal is to determine the wavenumber at which the wavelet function is maximum in the frequency domain. He determined that, in the case of the DoG wavelet, the equivalence between the wavelet scale and the wavelet scaling factor is $k=\sqrt(m)/l_{\textrm{F}}$, so that the scaling factor becomes $l_{\textrm{F}} = \sqrt{m}\cdot l\cdot(2\pi)^{-1}$.

In other words, for this analysis, the wavelet scaling factor is not chosen to correspond to the standard deviation of the initial Gaussian, but it is chosen to correspond instead to the Fourier wavenumber $k$ that is sampled by the bandpass filter in the Fourier domain. This definition allows a better comparison between the wavelet power spectrum and the classical Fourier power spectrum.

\section{Tests on simulated data}\label{sec:test}

The formalism presented in the previous section shares similarities with the $\Delta$-variance introduced by \citet{1998A&A...336..697S}. This method has been successfully applied in several studies in order to characterise structures at multiple scales induced by turbulence in molecular clouds \citep{2001A&A...366..636B, 2008A&A...485..917O, 2008A&A...485..719O, 2014A&A...568A..98A}. The $\Delta$-variance is defined as a measure of the amount of structure at a given scale $l$ in a map. Its definition is similar to the energy spectrum defined in equations (\ref{eq:wavelet_energy}) and (\ref{eq:Fourier_wavelet}), except that the convolved filter $\psi(\bmath{x})$ is isotropic. As mentioned by \citet{2008A&A...485..917O}, the main advantage of the $\Delta$-variance method comes from its smooth filter shape which ensures a robust angular average of the signal and a lower sensitivity to singular variations and finite map size effects. Similarly to the work of \citet{2001A&A...366..636B} which tested the influence of telescope beam and finite map sizes on the $\Delta$-variance, this section tests those effects using the anisotropic DoG wavelet.

\begin{figure*}
\centering
\includegraphics[]{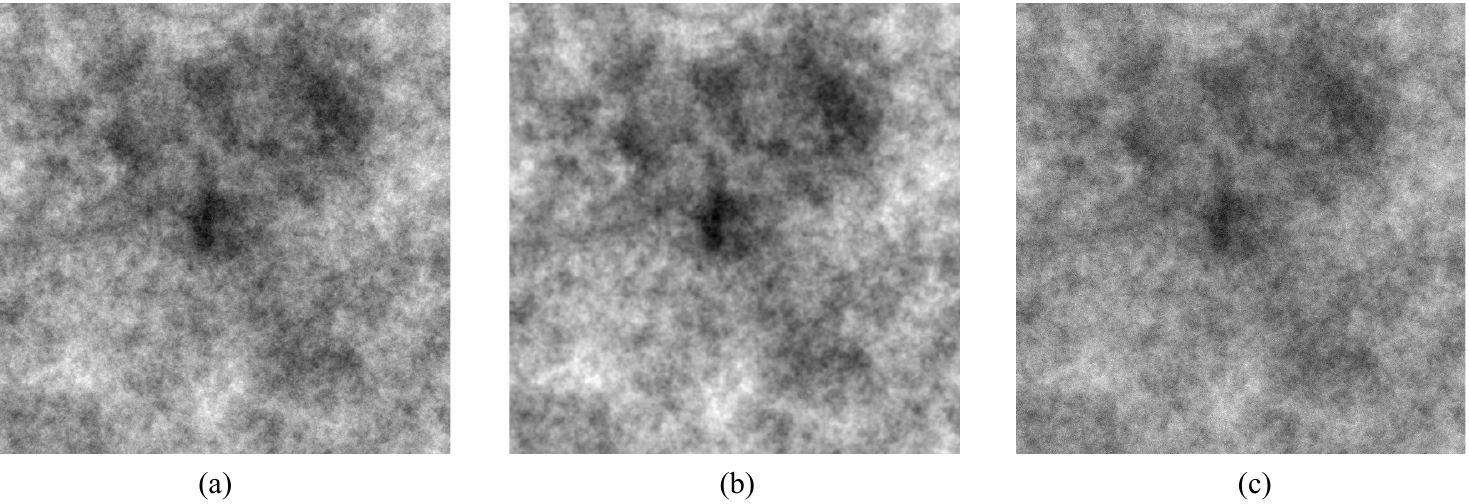}
\caption{The Grf simulation of $1024\times1024$ pixels of Stokes $Q$ map with $\gamma=2.5$: (a) the original Grf with a mean pixel value of zero and a standard deviation of 1.0, (b) the same field convolved with a Gaussian filter with a standard deviation of 2 pixels and (c) the original field with random noise, with an rms of 0.5, added.}
\label{fig:Qfbm2p5}
\end{figure*}

\begin{figure*}
\centering
\includegraphics[]{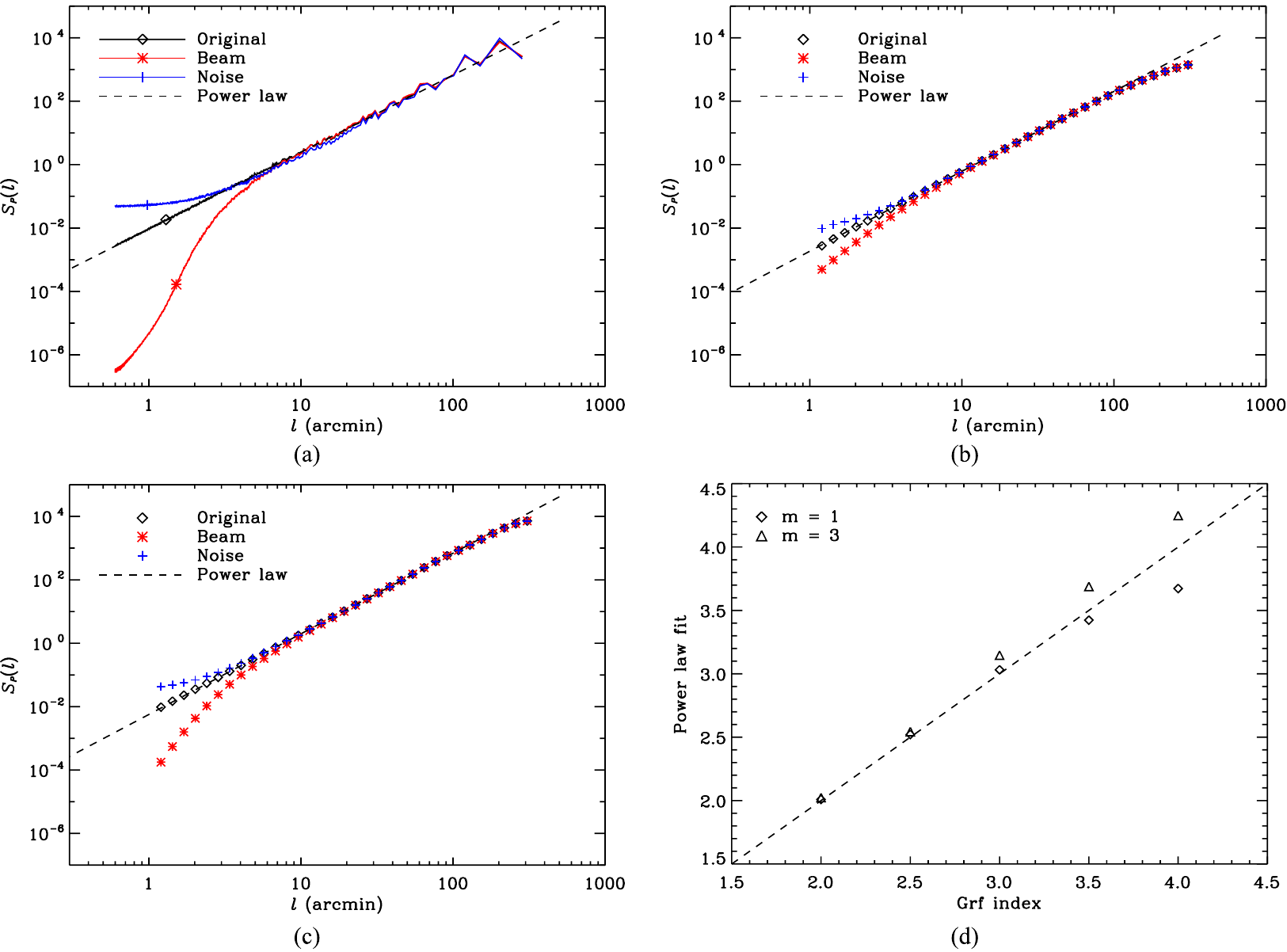}
\caption{Power spectra analysis of Grfs simulated Stokes $Q$ and $U$ maps: (a) The Fourier power spectrum of $|\bmath{P}|$ of the original image (diamond), the image convolved with a Gaussian beam (star) and the image with added noise (plus). (b) The wavelet power spectrum of the three same images using the first order wavelet. (c) The wavelet power spectrum of the three same images using the third order wavelet. (d) shows the values of the fitted power laws to the wavelet power spectra (for $4 < l < 50$ arcmin), for five different power law indices of the original Grfs. Diamonds represent power laws measured with the first order Dog wavelet ($m=1$) and triangles represent power laws measured with the third order Dog wavelet ($m=3$).}
\label{fig:fbm_power_spec_wavelet}
\end{figure*}

The robustness of the wavelet power spectrum calculation is tested on two Gaussian random field (Grf) simulations of $1024\times1024$ pixels for both Stokes $Q$ and $U$ maps. Those images are produced by applying a power law as a function of scale to the squared amplitude of a random-phase map. Similarities between Grfs and interstellar structures was pointed out by \citet{1998A&A...336..697S} in their study of fractal properties of molecular clouds. A Grf simulation with a power law of $\gamma=2.5$ representing Stokes $Q$ maps is displayed in Fig. \ref{fig:Qfbm2p5}. The original Grf is displayed in Fig. \ref{fig:Qfbm2p5}(a), Fig. \ref{fig:Qfbm2p5}(b) shows the same field convolved with a Gaussian filter having a standard deviation of 2 pixels and Fig. \ref{fig:Qfbm2p5}(c) shows the original field added with a random noise having a $\sigma_{\textrm{rms}}$ of 0.5. The original field has a mean pixel value of zero and a standard deviation of 1.0. The Fourier power spectra of $|\bmath{P}|$ for the three fields are shown in Fig. \ref{fig:fbm_power_spec_wavelet} (a). They are calculated on $2048\times2048$  extended maps with zero-padding and an apodised interface \footnote{The apodisation consists of the multiplication of a taper function, in this case the negative slope of a cosine, which smooths the sharp edges in an image.} between the extension and the image on 5 per cent of the border of the original image. To avoid spurious power at smaller scales caused by edges of the image, the mean pixel value of images must be subtracted before the apodisation. In Fig. \ref{fig:fbm_power_spec_wavelet} (a), the spectra are produced by averaging the squared amplitude of complex Fourier coefficients over different annuli in the $u$--$v$ plane. Figure \ref{fig:fbm_power_spec_wavelet} (a) shows that, at small scales, the telescope beam and noise induce a significant departure from the power law. 

\begin{figure}
\centering
\includegraphics[]{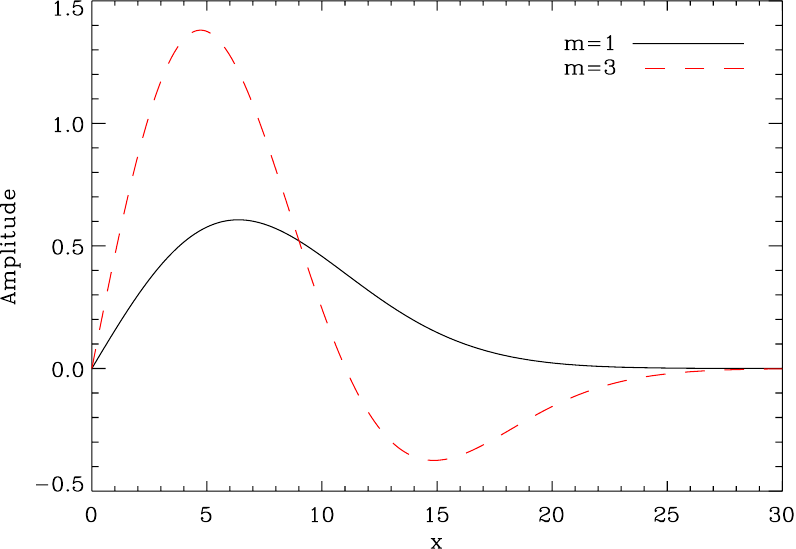}
\caption{The positive part of the first and the third derivative of a Gaussian centred at zero, respectively $m=1$ and 3 in equation (\ref{eq:mother_DoG}).\label{fig:DoG_1d}}
\end{figure}

The wavelet power spectra of $|\nabla \tilde{\bmath{P}}|$ calculated following equations (\ref{eq:scaled_gradient_P}) and (\ref{eq:wavelet_power}) are displayed in Fig. \ref{fig:fbm_power_spec_wavelet} (b) and (c). The calculation is done on the same extended maps as for the Fourier power spectra. Equidistant values of $l$ in logarithmic scale are chosen starting from 4 to 1024 pixels with an interval of $2^{0.25}$ pixels. The same pixel resolution as the CGPS data has been assigned to the Grfs. Wavelet power spectra on figures \ref{fig:fbm_power_spec_wavelet} (b) and (c) are respectively associated with the first and the third derivative of a Gaussian, i.e. $m=1$ and 3 in equation (\ref{eq:mother_DoG}). As noticed by \citet{Arneodo2000} and \citet{2006ApJS..165..512K}, the robustness of the results can be tested using the DoG wavelet by repeating the analysis with a wavelet of a higher order. The first and the third derivative of a Gaussian have respectively one and three vanishing moments. A wavelet with more vanishing moments can represent more complex functions. For example, the wavelet transform of a polynomial function of degree $n$ will be equal to zero, if a wavelet has vanishing moments up to the order $m \ge n$ \citep{2010tah..book.....P}. Consequently, repeating the analysis with a wavelet of a higher order can confirm that the scaling behaviour of the wavelet transform is not dominated by the order of the analysing wavelet and can also highlight the effect of polynomial distributions changing the self-similar geometry of the data. For both orders of the DoG wavelet, a power law of $\gamma=2.5$ is fitted for $4 < l < 50$ arcmin on the wavelet power spectra of the original Grfs. The wavelet with more vanishing moments is significantly more sensitive to the beam smoothing effect. The third order wavelet is also more affected by the noise, but less than by the beam convolution. It is important to note that the noise wavelet power spectrum with the first and the third order DoG wavelet has a flat power law, i.e. $\gamma=0$ instead of $\gamma=2$ as with the $\Delta$-variance. This difference comes from the normalisation choice discussed previously in section \ref{sec:wav_pow}. The third order wavelet is also less affected by the edge effect at larger scales than the first order wavelet. As shown in Fig. \ref{fig:DoG_1d}, since wavelet functions decay as $\bmath{x}^{-n}$, where $n$ is the order of the wavelet, the third order wavelet has a better localisation in the spatial domain than the first order. For this reason, the third order wavelet is less affected by the zero-padding which decreases the power of large-scale structures.

Figure \ref{fig:fbm_power_spec_wavelet} (d) shows the values of the fitted power laws to the wavelet power spectra between $4 < l < 50$ arcmin, for five different power law indices of the original Grfs. For the first order wavelet analysis, an underestimation of the spectral index is measured for $\gamma > 3$. This effect was also noticed by \citet{2001A&A...366..636B} for the $\Delta$-variance analysis and was attributed to the fact that edge effects are significant for maps covering only a fraction of the spatially extended emission. This statement is true for steep spectral index, where large-scale structures dominate. However, an overestimation of the spectral index is measured for $\gamma \ge 3$ using the third order wavelet, even if edge effects are less important for this wavelet. In that case the overestimation of the spectral index can be attributed to the lower resolution of this wavelet in the spectral domain.

As for the $\Delta$-variance, our wavelet power spectrum can satisfactorily recover the power law index of the fractal simulations for $2.0 \le \gamma \le 3.0$. According to their statistical properties, Grfs have the same power law index in every direction. Following this property, the calculation of the power spectrum of $|\nabla\bmath{P}|$ using equations \ref{eq:scaled_gradient_P}, \ref{eq:Stokes_amplitude} and \ref{eq:wavelet_power} can recover the power law index of individual Stokes $Q$ and $U$ simulated maps. Because of the normalisation choice discussed previously in section \ref{sec:wav_pow}, the slope of the wavelet power spectrum of $|\nabla\tilde{\bmath{P}}|$ is equal to the power law of the Grfs. This is similar to the $\Delta$-variance where the slope $\alpha$ is related to the power law $\gamma$ following the relation $\alpha=|\gamma|-2$. Real Stokes $Q$ and $U$ data are spatially correlated but are not assumed to have exactly the same power law index. Intervening polarised emission and faraday rotation along the line-of-sight should induce spatial correlation between $Q$ and $U$ maps and should also modify the measured power law of the wavelet power spectrum of $|\nabla \tilde{P}(l,\bmath{x})|$. Consequently, the wavelet power spectrum of $|\nabla \tilde{P}(l,\bmath{x})|$ is a unique measure of the variations of the polarisation vector $\bmath{P}$ in the $Q$--$U$ plane as a function of the angular scale and should not be directly compared with the Fourier power spectrum of $|\bmath{P}|$ or of Stokes $Q$ and $U$ maps alone.

\section{Application to CGPS data}\label{sec:results}

The wavelet analysis technique described in Section \ref{sec:DoG} was applied to the CGPS field shown in Fig. \ref{fig:MM12_MN12_P} for both Stokes parameters $Q$ and $U$. Each map was extended to the closest power of 2 - in this case, $2048 \times 2048$, with zero-padding pixels and apodised on 5 per cent of the border of the original image. Angular scales $l$ are chosen following the same rules as for the Grf simulations.

\begin{figure*}
\centering
\includegraphics[]{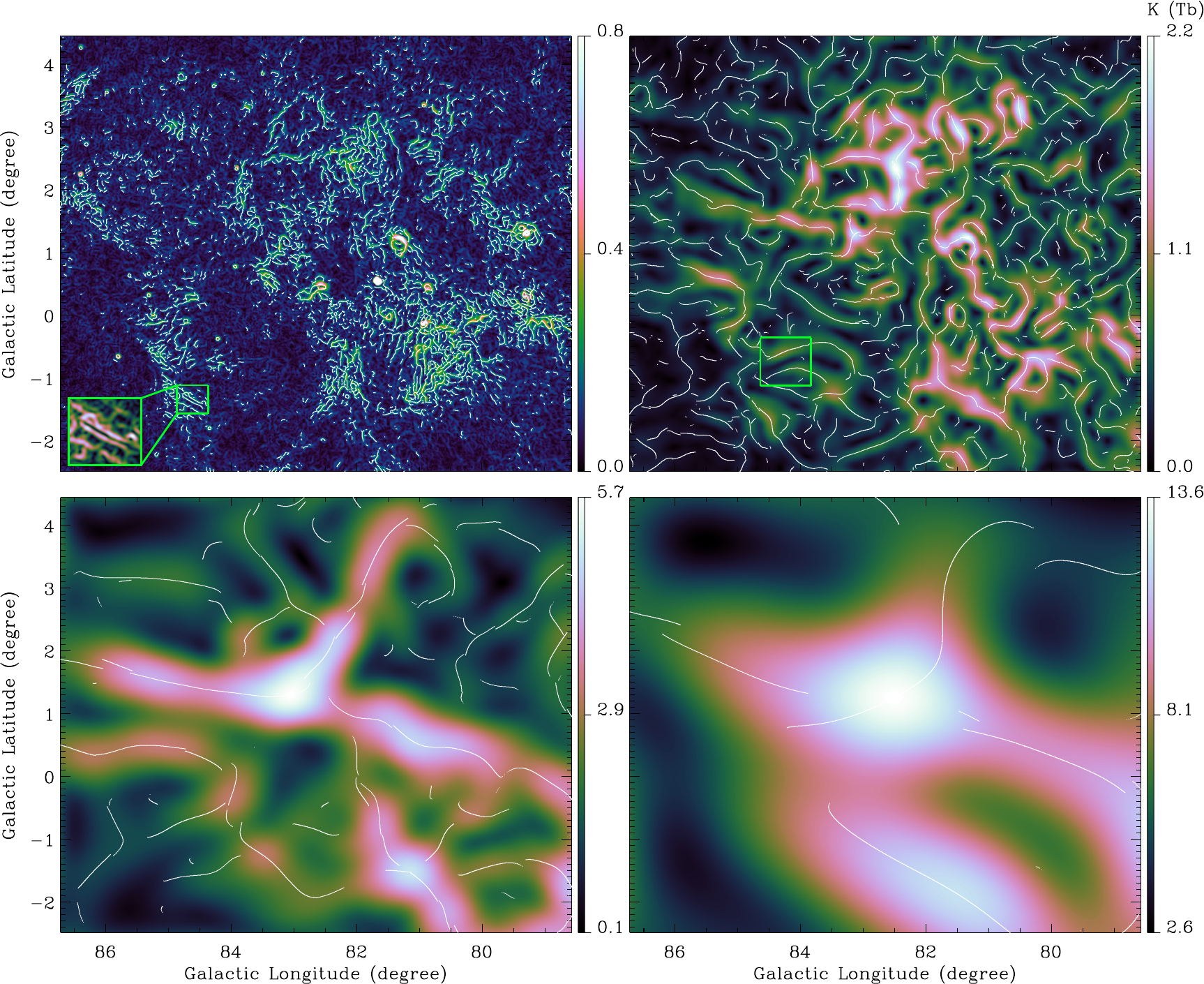}
\caption{From left to right: in ``cubehelix'' colour scale \citep{2011BASI...39..289G} are the $|\nabla \tilde{P}(l,\bmath{x})|$ values for the same field as in Fig. \ref{fig:MM12_MN12_P} at four different scales $l$= 9.6, 45.7, 153.6 and 434.4 arcmin. White lines represents maxima chains (see section \ref{subsec:maxima_chains}) corresponding to the scale.}
\label{fig:chains_scales}
\end{figure*}

\begin{figure*}
\centering
\includegraphics[]{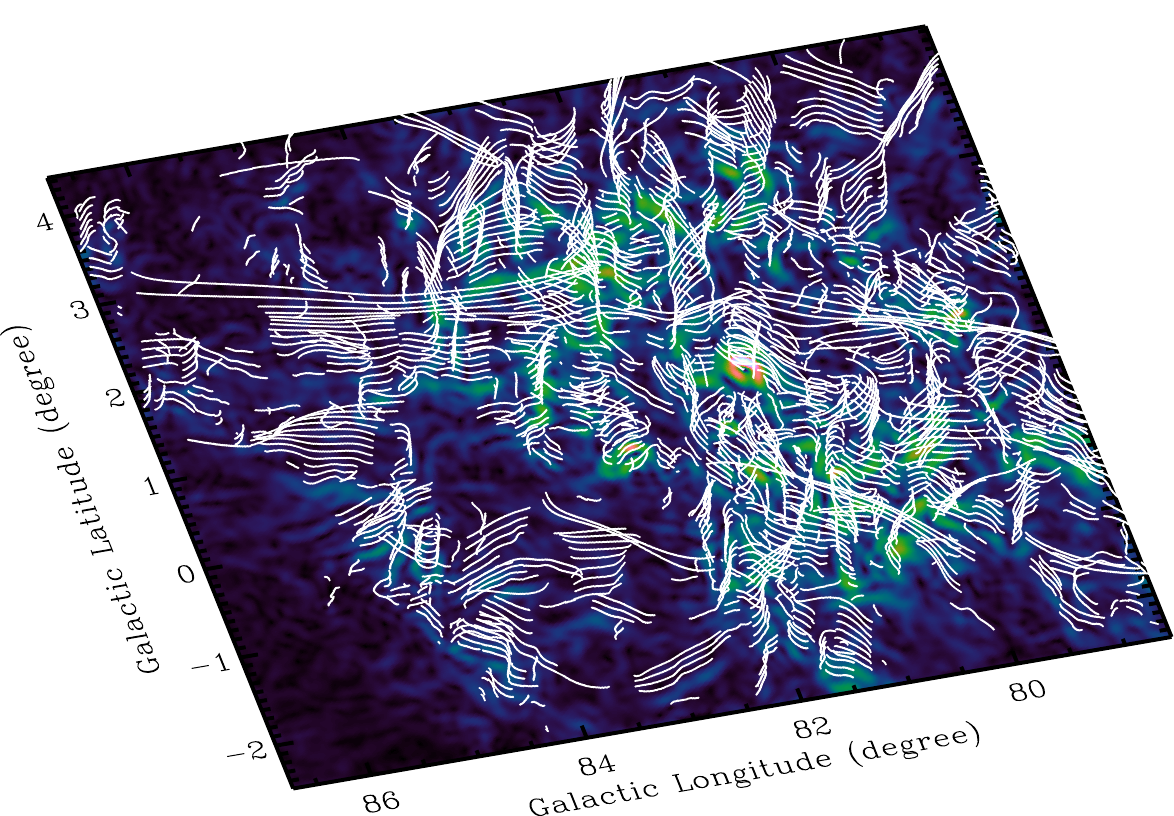}
\caption{The superposition of maxima chains from scale $l=$22.8 to 258.3 arcmin over the map of $|\nabla \tilde{\bmath{P}}|$ at $l=$ 22.8 arcmin. They represent a subset maxima chains for which the maximum value along the chain is part of outliers separated with the scale-wise CVE algorithm.} 
\label{fig:coherent_network}
\end{figure*}

The amplitude of the gradient of $\bmath{P}$ for the CGPS field at four different scales, overplotted with maxima chains, is displayed in Fig. \ref{fig:chains_scales}. For clarity, maxima chains of the smallest scale, $l=9.6$ arcmin, are only displayed for pixel values above $0.15$ K ($\approx 3\sigma_{\textrm{rms}}$). Each wavelet transform in Fig. \ref{fig:chains_scales} shows very different filamentary structures. The complex network of filaments at smaller scales is replaced by a more extended network of filaments on larger scales. The general pattern of the lower scale is sometimes preserved and sometimes not. Particularly, some features described as ``double jump'' profiles by  \citet{2012ApJ...749..145B} (see green boxes in left and right upper panels of Fig. \ref{fig:chains_scales}) appear only for a small range of scales. Such features are associated with the derivative of a delta function that can result from interactions of strong shocks. On the other hand, other subsets of the network persist over multiple scales and create a subset of ``coherent'' structures across the field. An example of a ``coherent'' subset network is displayed in Fig. \ref{fig:coherent_network}. The subset is selected using an iterative algorithm called the scale-wise Coherent Vorticity Extraction (CVE) (see \citet{2012PhyD..241..186N} and \citet{2014MNRAS.440.2726R} for a detailed description). As a function of scale $l$, this algorithm converges to an optimal threshold value to separate outliers, i.e. non-Gaussianities\footnote{By construction, following eq. \ref{eq:gradient_P}, the distribution of $|\nabla\tilde{P}(l,\bmath{x})|$ cannot respect a perfect Gaussian distribution, however the terminology Gaussian and non-Gaussian are used to describe respectively the symmetrical part and the tail of the distribution.}, from randomly distributed wavelet coefficients of $|\nabla\tilde{P}(l,\bmath{x})|$. Figure \ref{fig:coherent_network} shows maxima chains for which the maximum value along the chain is part of the separated outliers.

\begin{figure}
\centering
\includegraphics[]{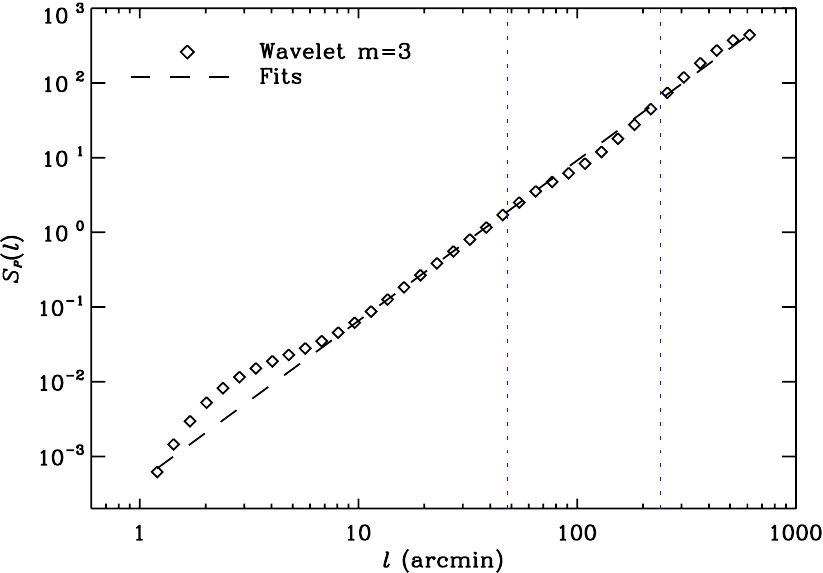}
\caption{The wavelet power spectrum of $|\nabla\tilde{P}(l,\bmath{x})|$ for the CGPS field for all coefficients (black diamonds).}
\label{fig:power_spec}
\end{figure}

\begin{figure}
\centering
\includegraphics[]{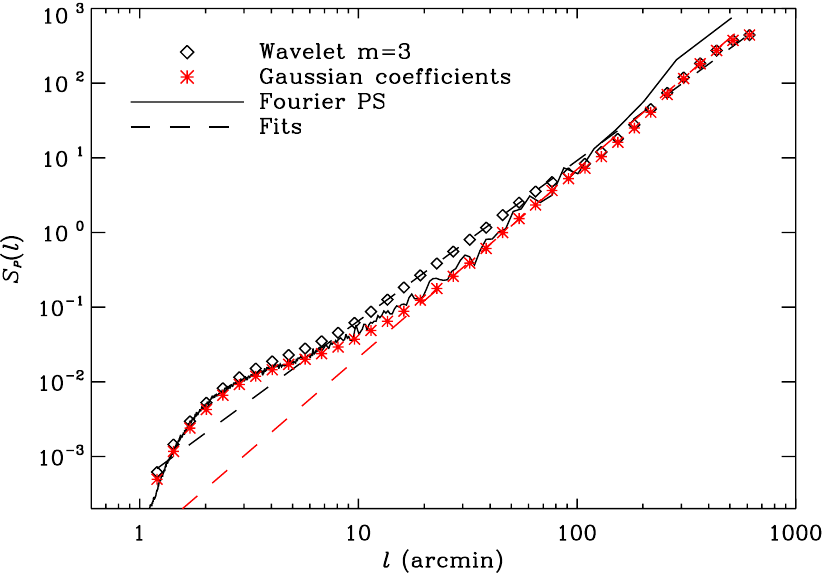}
\caption{The wavelet power spectrum of $|\nabla\tilde{P}(l,\bmath{x})|$ for the CGPS field for all coefficients (black diamonds) and for the Gaussian part of the distribution (red stars). The solid line represents the Fourier power spectrum of $|\mathbf{P}|$.}
\label{fig:power_spec_Gauss}
\end{figure}

The wavelet power spectrum calculated according to equation (\ref{eq:wavelet_power}) is shown in black diamonds in Fig. \ref{fig:power_spec} and \ref{fig:power_spec_Gauss}. Wavelet coefficients $|\nabla\tilde{P}(l,\bmath{x})|$ are calculated using the third order wavelet in order to highlight effects produced by noise and edges on the map (see section \ref{sec:test}). The flattening and the power drop caused by the Gaussian apodisation function applied to the DRAO Synthesis Telescope data \citep{2010A&A...520A..80L, 2014ApJ...787...34S} clearly appears in the power spectrum between $1 \lesssim l \lesssim 6$ arcmin. A flattening of the spectrum is also produced by the finite size of the map around 600 arcmin. The power spectrum shows a power law behaviour between $10 \lesssim l \lesssim 80$ arcmin followed by a small drop of power between $80 \lesssim l \lesssim 300$ arcmin. This drop corresponds well to the overlap in the $u$--$v$ plane between data from the Effelsberg 100--m telescope and the DRAO 26--m telescope, which is between baselines of 3 to 15 m \citep{2010A&A...520A..80L}. The corresponding angular sizes, 48 to 240 arcmin, are indicated by dotted lines in Fig. \ref{fig:power_spec}. The 26--m data were initially undersampled and gaps where no 26--m data were available were filled with smoothed Effelsberg data. This undersampling, and the process applied to correct the data, could have produced an underestimation of the power over that range of scales. The average calibration ratio between the two datasets is $0.96 \pm 0.01$ (26--m/100--m). This small factor could explain the drop in power seen in the wavelet power spectrum of the CGPS field. For this reason, a power law is fitted only between 10 to 60 arcmin following the relation $S_P(l)=S_0 \cdot l^{\gamma}$, where the fitted values of parameters $S_0$ and $\gamma$ are $(4.69\pm0.03)\times 10^{-4}$ and $2.15\pm0.01$ respectively.

\begin{figure}
\centering
\includegraphics[]{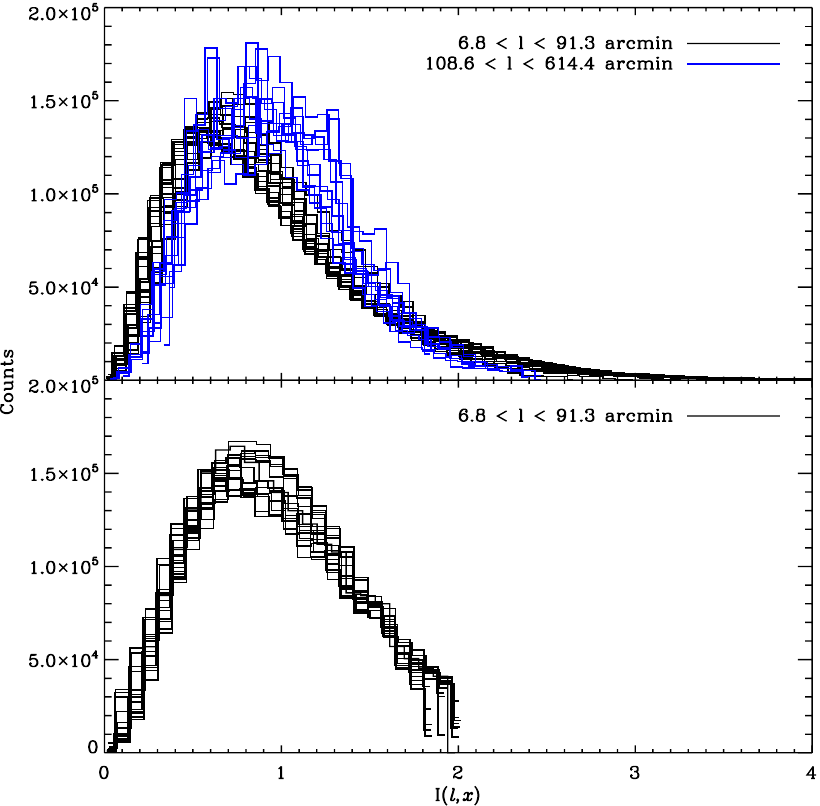}
\caption{Normalised distributions of wavelet coefficients of $|\nabla\tilde{P}(l,\bmath{x})|$ for the CGPS field for all coefficients (top panel). The black lines represent scales between 6.8 and 91.3 arcmin and the blue lines present scales between 108.6 and 614.4 arcmin. The lower panel shows the Gaussian part of the distribution for scales between 6.8 and 91.3 arcmin.}
\label{fig:coeff_distributions}
\end{figure}

The wavelet power spectrum has also been calculated using only the Gaussian coefficients of $|\nabla\tilde{P}(l,\bmath{x})|$ separated by the scale-wise CVE. This Gaussian power spectrum is represented by the red stars in Fig. \ref{fig:power_spec_Gauss}. Both distributions, for all coefficients and for the separated Gaussian part, are plotted with the black lines in Fig. \ref{fig:coeff_distributions}. The distributions are normalised following the definition: 

\begin{equation}
I(l,\bmath{x})=\frac{|\nabla\tilde{P}(l,\bmath{x})|}{\langle |\nabla\tilde{P}(l,\bmath{x})| \rangle_{\bmath{x}}},
\label{eq:intermittence}
\end{equation}

\noindent where $\langle\rangle_{\bmath{x}}$ is the average operator over all $\bmath{x}$. The normalised distributions for all coefficients between scales of 6.8 to 91.3 arcmin on the top panel of Fig. \ref{fig:coeff_distributions} (black lines) are lognormal and consequently, the average value of the coefficients does not accurately characterise the distribution. The separated Gaussian part shows a peak centred on $I(l,\bmath{x})\approx1$, which means that the average value of coefficients is more representative of the general tendency of the distribution. The fitted parameters for the Gaussian power spectrum are $(6.4\pm0.5)\times 10^{-5}$ and $2.52\pm0.02$ for $S_0$ and $\gamma$ respectively for scales between 20 to 60 arcmin. Scales $108.6 \lesssim l < 614.4$ arcmin do not respect the self-similarity of small-scale distributions. However, a clear separation between two different behaviours as a function of scale is hard to establish and the transition from lognormal to Gaussian distributions might also be continuous.

\section{Discussion}\label{sec:discussion}

The new method proposed in this paper allows us to extend to multiple scales the study of structures produced by the calculation of $|\nabla\bmath{P}|$. As shown in Fig. \ref{fig:chains_scales}, filamentary structures in $|\nabla\tilde{P}(l,\bmath{x})|$ are highly dependant on the angular scale or on the instrumental resolution with which the polarised signal is observed. Furthermore, the wavelet power spectrum analysis of $|\nabla\tilde{P}(l,\bmath{x})|$ allows us to identify scales where the signal to noise ratio or the beam signature becomes important and causes a flattening of the power law behaviour. Consequently, studies of the gradient of $\bmath{P}$ applied only to the smallest spacial scales, without or with little smoothing of original data, should be aware that a significant amount of the structure seen at lower intensity in $|\nabla\bmath{P}|$ images may be associated with noise.

Turbulence is expected to be one of the major processes responsible for fluctuations on multiple scales in the ISM. The power law measured across a large range of scales ($10 \lesssim l \lesssim 60$ arcmin), could be associated with the presence of turbulence in the magnetic field. The self-similarity of wavelet coefficient distributions plotted in Fig. \ref{fig:coeff_distributions} is another indication that turbulence can play a major role in fluctuations seen in the magnetic field and/or the electron density over the same range of scales. Although, as demonstrated with magnetohydrodynamic (MHD) simulations by \citet{2011Natur.478..214G} and \citet{2012ApJ...749..145B}, various types of turbulent environment, e.g. subsonic or supersonic turbulence with different Mach numbers, can create various type of structures in $|\nabla\bmath{P}|$ maps. As expected for a large field localised in the Galactic plane, the filament network of $|\nabla\bmath{P}|$ displays a large range of intensities and different types of discontinuity associated with different types of fluctuation in the magnetic field and/or the electron density. We see in Fig. \ref{fig:coeff_distributions} that even if the wavelet coefficient distributions are self-similar for a given range of scales, the power is not randomly distributed. The tails of the lognormal distributions shown in the upper panel are associated with coefficients that contribute more to the average power at a given scale than randomly distributed coefficients. This excess has a significant impact on the measured power law. The Gaussian part of the distributions for the self-similar range of scales possess a steeper power law ($\gamma \approx 2.5$) than the original distribution ($\gamma \approx 2.1$). The spatial distribution of non-Gaussianities displayed for a small range of scales in Fig. \ref{fig:coherent_network} also shows that these structures tend to create a coherent subset of filaments correlated across several scales. These coherent structures can have different origins. \citet{2012ApJ...749..145B} showed that moments of $|\nabla\bmath{P}|$, i.e. mean, variance, skewness and kurtosis, are correlated with the Mach number of MHD simulations. Higher Mach numbers create more asymmetric distributions which have tails at high intensity. Higher intensity structures in $|\nabla\bmath{P}|$ associated with those tails are caused by sharp changes of the polarisation vector $\bmath{P}$ that can be produced by compressive shocks in a supersonic turbulent medium. In dense regions, the magnetic field lines are frozen into the ionised gas and compressive shocks will amplify the magnetic field intensity. Under these conditions, the magnetic field intensity is correlated with the electron density and creates higher intensity structures seen in $|\nabla\bmath{P}|$ \citep{2012ApJ...749..145B}. However, subsonic turbulence induces no compressive motion and in that case fluctuations traced by $|\nabla\bmath{P}|$ are dominated by random fluctuations in the gradient of the magnetic field.

It is interesting to note in Fig. \ref{fig:power_spec_Gauss} that the power spectrum associated with the Gaussian part of the wavelet coefficient distribution corresponds well to the Fourier power spectrum of $|\mathbf{P}|$ for $l \lesssim 100$ arcmin. The selection of the Gaussian part of the distribution is dependent on a parameter in the scale-wise CVE algorithm which controls how restrictive the definition of an outlier is \citep{2012PhyD..241..186N}. Nonetheless, the choice of this parameter is based on the symmetry of the separated ``random'' distributions only. By comparing the total wavelet power spectrum of $|\nabla\tilde{P}|$ with the classical Fourier power spectrum of $|\mathbf{P}|$, we see that the excess of power measured at intermediate scales might be associated with non-Gaussianities identified from the wavelet analysis. The amplitude of $\bmath{P}$ and the amplitude of the gradient of $\bmath{P}$ are two different tracers and they are not expected to produce the same power spectrum. The amplitude of the gradient of $\bmath{P}$ traces changes in the polarisation vector more strongly than the amplitude of $\bmath{P}$ alone. Consequently, the excess of power measured at intermediate scales, partly displayed in Fig. \ref{fig:coherent_network}, might be related to the expected correlation between the magnetic field intensity and the electron density produced by compressive shocks. This measured excess of power suggests that the power spectrum of $|\nabla\tilde{P}|$ may trace fluctuations in the electron density as well as fluctuations caused by the Faraday rotation of polarised emission coming from a source localised behind the compressed magnetic field lines, whereas the power spectrum of $|\bmath{P}|$ traces only fluctuations of the electron density.

The power law index of the Gaussian power spectrum is shallower than that expected from a 3D Kolmogorov-like power spectrum ($\gamma=3.66$) for a subsonic incompressible turbulent medium, however it is close to the index measured by \citet{2002A&A...387...82G} for the polarised intensity $\bmath{P}$ at 2.4 GHz from the Southern Galactic plane ($\gamma= 2.37 \pm 0.21$). In the case of the non-Gaussian subset, many other physical processes can produce sharp changes in polarised data, such as outflows from massive stars and supernovae. We noted in section \ref{sec:results} that features described as ``double jumps'' by \citet{2012ApJ...749..145B} are visible at different scales in the CGPS field. One of the two features framed by a green rectangle (the upper right panel at $l=45.7$ arcmin) corresponds to the location of the supernova remnant SNR G84.2-0.8, which is clearly seen in the total intensity map. The ``double jump'' feature, which is itself part of the non-Gaussian subset, may be related to shocks produced by the supernova.

On scales of $109 \lesssim l \lesssim 614$ arcmin, we can see in Fig. \ref{fig:coeff_distributions} (blue lines) that distributions become more symmetrical on larger scales. Since the distributions are normalised, it is reasonable to consider that the undersampling correction applied to the 26--m Telescope data, which dominate on scales larger than $\sim 80$ arcmin, does not influence or bias distributions in a strong manner. The fitted power law with the Gaussian part of the distributions at lower scales could also be consistent with scales dominated by the DRAO 26--m, which could mean that non-Gaussian coefficients tend to appear at smaller scales. According to \citet{2012ApJ...749..145B}, the random distribution of Gaussian coefficients could be induced by subsonic turbulent fluctuations in electron density and magnetic field. On the other hand, non-Gaussian coefficients may be associated with other physical processes, such as supersonic turbulence, gravitational collapses, massive star outflows and supernovae. However, a clear separation between lognormal distributions at smaller scales and Gaussian distributions at larger scales is hard to confirm. The lack of non-Gaussian contributions at larger scales could also be associated with the smaller number of statistically independent coefficients. A similar analysis realised on an extended range of scales could confirm if a real separation exists between distributions at smaller and larger scales.

\section{Conclusion}\label{sec:conclusion}

We extend the calculation of $|\nabla\bmath{P}|$ to multiple scales using a wavelet analysis formalism. The new technique shares similarities with the $\Delta$-variance and the WTMM techniques used to characterise respectively the turbulence in molecular clouds and the multifractal nature of a surface or a medium. This approach can overcome the limitation of previous analyses which were only sensitive to the smallest scales. We show that fluctuations traced by $|\nabla\bmath{P}|$ exist at larger scales on data completed with lower spatial frequencies. Using the wavelet formalism, it is possible to measure the power spectrum of $|\nabla\tilde{P}(l,\bmath{x})|$ and evaluate the scaling behaviour of variations of the polarisation vector $\bmath{P}$ in the $Q$--$U$ plane. The scaling behaviour fallows a power law with $\gamma \approx 2.1$. We measure a small drop in the spectrum between $80 \lesssim l \lesssim 300$ arcmin. This drop corresponds well to the overlap in the $u$--$v$ plane between the Effelsberg 100--m telescope and the DRAO 26--m telescope data. The undersampling presents in the 26--m telescope data has been identified as a source of unknown error in the data and could explain the measured drop of power. The wavelet analysis of $|\nabla\bmath{P}|$ also allows us to analyse the distribution of fluctuations in $\bmath{P}$ as a function of angular scale. Distributions show higher skewness at smaller scales than at larger scales. Separation of outliers contributing to the tails of the distributions allows us to measure the power spectrum for the symmetrical part of the distribution. This power spectrum possesses a steeper power law with $\gamma \approx 2.5$. The spatial distribution of some outliers are part of correlated structures across angular scales, which trace the sharpest changes in the polarisation vector $\bmath{P}$ in the field. Such higher intensity structures could be associated with compressive shocks of a supersonic turbulent medium. Future analysis applied over an extended range of scales and at higher Galactic latitude will provide a useful extension to the analysis presented here. Such analysis could confirm the appearance of a distinct type of fluctuation distribution at smaller scales as well as revealing the presence of high intensity structures at higher Galactic latitude or at higher angular scales.

\bibliographystyle{apj}
\bibliography{biblio}

\end{document}